\documentclass[12pt]{article}
\usepackage{amsmath}
\usepackage{amscd}
\usepackage{amssymb}
\usepackage{array}

\begin{document}
\rightline{CERN-TH/2001-377}

\newcommand{\id}{\relax{\rm 1\kern-.35em 1}}

\newcommand{\Z}{\mathbb{Z}}
\newcommand{\R}{\mathbb{R}}
\newcommand{\C}{\mathbb{C}}
\newcommand{\N}{\mathbb{N}}
\newcommand{\Ha}{\mathbb{H}}

\newcommand{\fiso}{\mathfrak{iso}}
\newcommand{\fso}{\mathfrak{so}}
\newcommand{\fosp}{\mathfrak{osp}}
\newcommand{\fsp}{\mathfrak{sp}}
\newcommand{\fsl}{\mathfrak{sl}}
\newcommand{\fsu}{\mathfrak{su}}
\newcommand{\fspin}{\mathfrak{spin}}
\newcommand{\ff}{\mathfrak{f}}
\newcommand{\fg}{\mathfrak{g}}
\newcommand{\fu}{\mathfrak{u}}
\newcommand{\fo}{\mathfrak{o}}
\newcommand{\fgl}{\mathfrak{gl}}
\newcommand{\fii}{\mathfrak{i}}
\newcommand{\fusp}{\mathfrak{usp}}

\newcommand{\g}{\mathcal{G}}
\newcommand{\s}{\mathcal{S}}
\newcommand{\ka}{\mathcal{K}}
\newcommand{\p}{\mathcal{P}}
\newcommand{\A}{\mathcal{A}}
\newcommand{\I}{\mathcal{I}}
\newcommand{\Ca}{\mathcal{C}}
\newcommand{\La}{\mathcal{L}}
\newcommand{\Qa}{\mathcal{Q}}

\vskip 1.5cm

  \centerline{\LARGE \bf Considerations on Super Poincar\'e Algebras and  }

  \smallskip

\centerline{\LARGE \bf  their Extensions to Simple Superalgebras }
 \vskip 3cm
\centerline{S. Ferrara$^\star$ and M. A. Lled\'o$^\dagger$.}

\vskip 1.5cm

\centerline{\it $^\star$ CERN, Theory Division, CH 1211 Geneva 23,
Switzerland, } \centerline{\it INFN, Laboratori Nazionali di
Frascati, Italy and } \centerline{\it Department of Physics and
Astronomy, University of California, Los Angeles, U.S.A.}

\medskip

\centerline{\it $^\dagger$  INFN, Sezione di Torino, Italy and }
\centerline{\it Dipartimento di Fisica, Politecnico di Torino,}
\centerline{\it Corso Duca degli Abruzzi 24, I-10129 Torino,
Italy}

\vskip 1cm

\begin{abstract}
We consider simple superalgebras which are a supersymmetric
extension of the spin algebra in the cases where the number of odd
generators does not exceed 64. All of them contain a super
Poincar\'e algebra as a contraction  and another as  a subalgebra.
Because of the contraction property, some of these algebras can be
interpreted as de Sitter or  anti de Sitter superalgebras.
However, the number of odd generators present in the contraction
is not always minimal due to the different splitting properties of
the spinor representations under a subalgebra.  We consider the
general case, with arbitrary dimension and signature, and examine
in detail particular  examples with physical implications in
dimensions $d=10$ and $d=4$.
\end{abstract}

\vfill\eject

\section{Introduction}

Super Poincar\'e algebras \cite{wz} are non semisimple
superalgebras  \cite{nrs,ka,bg}.  Their even part is the
Poincar\'e algebra (plus some extra generators that we will see
below) and their odd part carries one   or more (for $N$-extended
supersymmetry) spinor representations of the underlying Lorentz
algebra.

 A {\it spinor representation} of the Lorentz algebra is
an irreducible complex representation whose highest weights are
the fundamental weights corresponding to the right extreme nodes
in the Dynkin diagram. These are representations of the spin group
that do not descend to representations of the orthogonal
group.(For a review see Ref.\cite{st,ca}).
 We will call
{\it odd charges} or {\it spinor charges} the generators of the
odd part of the super Poincar\'e algebra. A reality condition must
be imposed on the spinor charges to obtain a real Lie
superalgebra.

Generically, we can write the anticommutator  of two spinor
charges as
\begin{equation}\{Q^I_\alpha,Q^J_\beta\}=\gamma^\mu_{\alpha\beta}p_\mu G^{IJ}
 +\sum\limits_{k} \gamma^{[\mu_1\cdots
\mu_k]}_{(\alpha \beta)}Z^{[IJ]}_{[\mu_1\cdots \mu_k]}.
\label{ccharges}\end{equation}  Here the indexes $\alpha$ run over
the spinor representation, and $I,J=1,\dots N$.
$Z^{[IJ]}_{[\mu_1\cdots \mu_k]}$ are even generators that are in a
antisymmetric tensorial representation (a representation on the
antisymmetric $n$th tensor product of the fundamental
representation space) of the Lorentz group and commute with the
translation generators.

In general there is a group G that acts on $Q^I$, which depends on
the particular properties of the spinor representations in
different signatures and dimensions. $G^{IJ}$ is then an invariant
tensor under this
 group and $Z^{IJ}$ is in the two fold (symmetric or
 antisymmetric) representation of such group. It is called the
 automorphism group because its action leaves invariant the Lie
 superbrackets of the Poincar\'e superalgebra (it acts trivially on the other generators).
 The
 symmetry properties  of $Z^{IJ}$ are the same than the symmetry
 properties  (in $\alpha,\beta$) of $\gamma^{[\mu_1\cdots
\mu_k]}_{(\alpha \beta)}$, which in turn depend only on the
space-time dimension modulo 8 (and not on the signature).

From the bracket (\ref{ccharges}), we see that  the even part of a
super Poincar\'e algebra has  an abelian ideal that contains the
spacetime translation generators and the $Z$ generators. Together
with the odd charges they form a central extension of the
supertranslation algebra. The physical interpretation of the $Z$
generators is related to p-brane charges \cite{du,gt}, that is, to
certain configurations of supergravity theories in which the
expectation value of the $Z$-charges is non zero.

Superconformal algebras are simple supersymmetric extensions of
the conformal algebra. In Ref.\cite{dflv} these extensions for
$N=1$ were studied for arbitrary dimension ($d=s+t)$ and signature
($\rho=s-t$) and in Ref.\cite{dfl} the case of extended
supersymmetry was treated. There are in general two superconformal
algebras, a maximal one which is always $\fosp(1|2Nn)$ ($2n$ is
the dimension of the spinor representation of the odd generators)
and a minimal one, whose bosonic part is the Spin(s,t)-algebra of
the spinor representation. Only in  dimensions $d=3,4,5,6$ is it
possible to find a simple superalgebra with a bosonic part which
factorizes as a direct sum of the orthogonal algebra $\fso(s,t)$
 plus a simple  R-symmetry algebra \cite{na}, as required
by the Coleman-Mandula theorem \cite{cm}. The odd part of the
superalgebra is a direct sum spinor representations  \cite{hls}.
For higher dimensions, the bosonic part is also the direct sum of
an R-symmetry subalgebra and a spacetime subalgebra, but the
spacetime subalgebra is enlarged with extra generators.

 In the web of connections between
string theory and M -theory \cite{wi}, or possible generalizations
as F-theory \cite{va} or S-theory \cite{ba}, it is natural to
investigate the role played by the simple superalgebra,  even in
the cases $d>6$ \cite{dflv,dfl,to,to2,ho,bvp}. As in the purely
bosonic case, the super Poincar\'e algebra with $Z$-charges can be
obtained from simple superalgebras in two different ways. One is
by contraction \cite{df}, the other as sub-superalgebra
\cite{to,to2}.

Let spacetime be  a manifold of dimension $d$ with an (indefinite)
metric of signature $(s,t)$. The Poincar\'e group acts on flat
spacetime, $\R^d\simeq \break{\rm ISO}(s,t)/{\rm SO}(s,t)$. Its
Lie algebra, $\fiso(s,t)$, is a subalgebra the conformal algebra
of $\R^d$, the simple algebra $\fso(s+1,t+1)$ \cite{affs}. Other
possible backgrounds are the symmetric spaces ${\rm
SO}(s+1,t)/{\rm SO}(s,t)$, with signature $(s,t)$. There is a
contraction of the isometry algebra $\fso(s+1,t)$ which gives the
Poincar\'e algebra $\fiso(s,t)$. Interchanging the roles of $s$
and $t$ we have a different contraction, from $\fso(s,t+1)$. For
physical signature, the two spaces are  the de Sitter space ${\rm
SO}(d,1)/{\rm SO}(d-1,1)$ and the anti de Sitter one ${\rm
SO}(d-1,2)/{\rm SO}(d-1,1)$. Unitary multiplets of the anti de
Sitter superalgebra in dimension 11, $\fosp(1|32)$ where
investigated in Ref.\cite{gu}.

 The super
Poincar\'e algebra in dimension $d$ is a subalgebra of the
superconformal algebra in the same dimension, appearing with
different number extra generators in the abelian ideal, depending
whether one looks at the minimal or the maximal superconformal
algebra. Ref.\cite{we}
 is an attempt to formulate M-theory as a spontaneously
broken phase of its superconformal extension, where the symmetry
under the  superconformal algebra $\fosp(1|64)$ is broken to the
super Poincar\'e subalgebra with 2 and 5-brane charges.

Anti de Sitter and de Sitter superalgebras are  supersymmetric
extensions of $\fso(d-1,2)$ and  $\fso(d,1)$ respectively. They
play an important role in the framework of the AdS and dS/CFT
duality \cite{agmoo,wi2,str,hu2}.

 Also, it is possible to obtain the super Poincar\'e
algebra as a contraction of a simple superalgebra. We have   the
de Sitter and anti de Sitter superalgebras (simple extensions of
the de Sitter and anti de Sitter algebras respectively), although
not always the contraction of an $N=1$ super algebra gives an
$N=1$ super Poincar\'e algebra. In Ref.\cite{ho} the possibility
of using some superalgebra gauge theory which gives M-theory  as a
particular low energy configuration (contraction) is explored.

Simple superalgebras embedding ordinary spacetime supersymmetry
algebras are also relevant to explore how the theories depend upon
the signature of space-time and provide a clue on the existence of
supergravity theories with non lorentzian spacetime signature, as
conjectured in Ref.\cite{hu} on the basis of time like T duality,
with the M, M' and  M$^*$-theories in eleven dimensions.

\smallskip

The paper is organized as follows. In Section \ref{charges} we
enumerate all the (minimal) superconformal algebras with 64, 32,
16 and 8 spinor charges in dimensions $d=3,\dots 11$ and arbitrary
signature. We observe that the same superalgebra may be obtained
from spacetimes with different signatures $\rho, \rho'$ if they
are congruent mod 8, $\rho=\pm \rho'+8n$, which may suggest a
duality of the physical theories. In Section 3 the de Sitter and
anti de Sitter superalgebras in dimensions $d=3,\dots 12$ are
considered and their contractions to super Poincar\'e algebras are
studied. In Section 4 we consider physically interesting examples
in $d=4$ and $d=10$. In the Appendix we give some basic
definitions about Lie superalgebras.

\section{Super conformal algebras in diverse dimensions\label{charges}}

Superconformal algebras with up to 64 spinor real charges
correspond to different real forms of complex superalgebras whose
even part contains $\fso(s+1,t+1)$ and whose odd part is a direct
sum of spinor representations of the same algebra. A spinor in
dimension $d+2$, $d=s+t$, has complex dimension  $2^{(d+1)/2}$ for
$d$ odd and $2^{d/2}$ for $d$ even (chiral spinors). The dimension
of the real representation depends on  the reality condition (the
same than the complex in the real case, twice the complex
dimension in the quaternionic and complex case).  In even
dimension, when the superalgebra is of type $\fsl(m|n)$
($d+2=2,6$), it contains left and right spinors (non chiral
algebra) while if the algebra is of type $\fosp(m|n)$ then it is
chiral (in fact, the metric preserving condition halves the number
of odd generators with respect to the linear superalgebra). For
example, for $d=12$ the superconformal algebra is linear (or
unitary) and for $N=1$ it has already 128 charges. So the maximal
dimension that we can consider is $d=11$.

\paragraph{$d=11$.} For $\rho=1,7$ mod 8 we have $\fosp(1|64)$ with 64 odd charges. It
corresponds to spacetimes of type (10,1) (M-theory), (9,2)
(M$^*$-theory) and (6,5) (M'-theory) \cite{hu}.

\paragraph{$d=10$.}
For $\rho=0$ mod 8, we have $\fosp(2-q,q|32)$, ($q=0,1$) with 64
charges and  $\fosp(1|32)$ with 32 charges. They correspond to
spacetimes of type (5,5) and (9,1).

\smallskip

\noindent For $\rho=2,6$ mod 8  we have $\fosp(1|32,\C)$ with 64
odd charges and spacetimes of type (6,4), (10,0) and (8,2).

\smallskip

\noindent For $\rho=4$ we have $\fosp(2^*|16,16)$ with 64 charges
and spacetime (7,3).

These correspond to different forms of Type IIA, IIB and (1,0)
theories studied in Ref.\cite{hu}.

\paragraph{$d=9$.} For $\rho=1,7$ mod 8 we have $\fosp(2-q,q|32)$
($q=0,1$) with 64 charges  and  $\fosp(1|32)$ with 32 charges.
They correspond to   spacetimes of type (9,0), (5,4), (8,1).

\smallskip

\noindent For $\rho=3,5$ we have $\fosp(2^*|16,16)$ with 64
charges and spacetimes of type (6,3) and (7,2).

\paragraph{$d=8$.} For $\rho=0$ mod 8 we have $\fsl(16|2)$ with 64 charges
and  $\fsl(16|1)$ with 32 charges. They correspond to  spacetimes
of types (4,4) and (8,0).

\smallskip

\noindent For $\rho=2,6$ we have $\fsu(8,8|2-q,q)$ ($q=0,1$) with
64 charges and $\fsu(8,8|1)$ with 32 charges. They correspond to
 spacetimes of type (5,3) and (7,1).

\smallskip

\noindent For $\rho=4$ mod 8 we have $\fsu^*(16|2)$ with 64
charges and spacetime of type (6,2).

\paragraph{$d=7$.}For $\rho=1,7$ we have $\fosp(8,8|4)$ with 64
charges and $\fosp(8,8|2)$ with 32 charges. They correspond to
spacetimes of type (4,3) and (7,0).

\smallskip

\noindent  For $\rho=3,5$ we have $\fosp(16^*|4-2q,2q)$ ($q=0,1$)
with 64 charges and $\fosp(16^*|2)$ with 32 charges. They
correspond to  spacetimes of type (5,2) and (6,1).

\paragraph{$d=6$.} For $\rho=0$ we have $\fosp(4,4|8)$ with 64
charges, $\fosp(4,4|4)$ with 32 charges and $\fosp(4,4|2)$ with 16
charges. They correspond to spacetime of type (3,3).

\smallskip

\noindent For $\rho=2,6$ we have $\fosp(8|4,\C)_\R$ with 64
charges and $\fosp(8|2,\C)$ with 32 charges. They correspond to
spacetimes of type (6,0), (4,2).

\smallskip

\noindent For $\rho=4$ we have $\fosp(8^*|8-2q,2q)$ ($q=0,1,2$)
with 64 charges, $\fosp(8^*|4-2q,2q)$ ($q=0,1,2$) with 32 charges
and $\fosp(8^*|2)$ with 16 charges. They correspond to spacetime
of type (5,1).

 \paragraph{$d=5$.}For $\rho=1$ we have $\fosp(4,4|8)$  with 64 charges, $\fosp(4,4|4)$
 with 32 charges and $\fosp(4,4|2)$ with 16 charges. They
 correspond to
spacetime of type (3,2).

\smallskip

\noindent  For $\rho=3,5$ we have $\fosp(8^*|8-2q,2q)$ ($q=0,1,2$)
with 64 charges, $\fosp(8^*|4-2q,2q)$ ($q=0,1$) with 32 charges
and $\fosp(8^*|2)$ with 16 charges. They correspond to spacetimes
of type (5,0) and (4,1).

For $d=5$ there exists a smaller superalgebra, the exceptional
superalgebra $\ff_4^p$. The integer number $p$ denotes the real
form of the complex superalgebra $\ff_4$, which depends on the
signature of spacetime.  For $\rho=3, 5$ the even part of the
superalgebra is $\fspin(7-p,p)\oplus\fsu(2)$ with $p=2,1$. In
these signatures, the spinors are quaternionic (pseudoreal), so
there exists a pseudoconjugation in the spinor space, which
together with the pseudoconjugation in the fundamental of
$\fsl(2,\C)$ defining $\fsu(2)$ gives a conjugation defining the
real form of the superalgebra. For signatures $\rho=1,7$ we have
an even part $\fspin(7-p,p)\oplus\fsl(2,\R)$, with $p=3,0$. In
these cases the spinors are real, so the conjugation defining the
real form of the superalgebra is formed with the conjugation in
the spinor space and the conjugation in the fundamental of
$\fsl(2, \C)$ defining $\fsl(2,\R)$.  The superalgebra has 16
charges. (For more on conjugations, pseudoconjugations and real
forms see Ref.\cite{dflv}. See also Ref.\cite{vp}

\paragraph{$d=4$.}For $\rho=0$ we have $\fsl(4| 8)$ with 64
charges, $\fsl(4| 4)$ with 32 charges, $\fsl(4| 2)$ with 16
charges and $\fsl(4|1)$ with 8 charges. They correspond to
spacetime of type (2,2).

\smallskip

\noindent For $\rho=2$  we have $\fsu(2,2|8-q,q)$ ($q=0,\dots 4$)
with 64 charges, $\fsu(2,2|4-q,q)$ ($q=0,1,2$) with 32 charges,
$\fsu(2,2|2-q,q)$ ($q=0,1$) with 16 charges and $\fsu(2,2|1)$ with
8 charges. They correspond to
 spacetime of type (3,1).

\smallskip

\noindent For $\rho=4$ mod 8 we have $\fsu^*(4|8)$ with 64
charges, $\fsu^*(4|4)$ with 32 charges, $\fsu^*(4|2)$ with 16
charges
 and spacetime of type (4,0).

 \paragraph{$d=3$.} For $\rho=1$  we have $\fosp(16-q,q|4)$ ($q=0,\dots 8$) with 64
 charges, $\fosp(8-q,q|4)$ ($q=0,\dots 4$) with 32 charges, $\fosp(4-q,q|4)$
 ($q=0,1,2$) with 16 charges, $\fosp(2-q,q|4)$ ($q=0,1$) with 8
 charges and $\fosp(1|4)$ with 4 charges. They correspond to  spacetime of type (2,1).

\smallskip

\noindent  For $\rho=3$  we have $\fosp(16^*|2,2)$ with 64
charges, $\fosp(8^*|2,2)$ with 32 charges, $\fosp(4^*|2,2)$ with
16 charges and  $\fosp(2^*|2,2)$ with 8 charges. They correspond
to  spacetime of type (3,0).

\section{De Sitter and anti de Sitter superalgebras and their contractions}
We write the simple superalgebras that are extensions of de Sitter
($\fso(d,1)$) and anti de Sitter  ($\fso(d-1,2)$) algebras in
physical signature $(d-1,1)$.

\begin{table}[ht]
\begin{center}
\begin{tabular} {|c|l|c||l|c||c|}
\cline{1-6}  $d$&  de Sitter& odd  & anti de Sitter&odd &odd SP
\\ \cline{1-6}  3&$\fosp(N|2,\C)_\R$&$4N$
&$\fosp(N-q,q|2)$&$2N$& $2N$\\  4&$\fosp(N^*|2,2)$&$4N$&
$\fosp(N-q,q|4)$&$4N$& $4N$\\
 5&$\fsu^*(4|N)$&$8N$ &$\fsu(2,2|N-q,q)$&$8N$&$8N$\\
6&$\fosp(8^*|2N-2q,2q)$&$16N$& $\fosp(8^*|2N-2q,2q)$&$16N$&$8N$\\
7&$\fosp(8|2N,\C)$&$32N$& $\fosp(8^*|2N-2q,2q)$&$16N$&$16N$\\ 8&
$\fosp(8,8|2N)$&$32N$ & $\fosp(16^*|2N-2q,2q)$&$32N$&$16N$\\ 9&
$\fsl(16|N)$&$32N$ &$\fsu(8,8|N-q,q)$&$32N$&$16N$\\10&
$\fosp(N-q,q|32)$&$32N$&$\fosp(N-q,q|32)$&$32N$&$16N$\\
11&$\fosp(N|32,\C)_\R$&$64N$& $\fosp(N-q,q|32)$&$32N$&$32N$\\ 12&
$\fosp(N^*|32,32)$&$64N$ &$\fosp(N-q,q|64)$&$64N$&$64N$\\
\cline{1-6}
\end{tabular}
\caption{De Sitter  and anti de Sitter
superalgebras.}\label{desitter}
\end{center}
\end{table}

For $d=4,5,12$, the de Sitter superalgebra exists only with $N$
even.

In $d=6$ and $d=10$ the de Sitter and anti de Sitter superalgebras
coincide. This is because the signature  $\pm \rho$  of the  the
$(d+1)$-dimensional spaces (where the de Sitter or anti de Sitter
algebras are linearly realized) are congruent modulo 8 (3 and 5
for $d=6$, 1 and 7 for $d=10$). For anti de Sitter we have that
the superalgebras of $d=6,7$ and $d=10,11$ coincide. For $d=6$
there exists a smaller subalgebra, the exceptional superalgebra
$\ff_4$ with two real forms, $\ff_4^1$ and $\ff_4^2$ whose bosonic
parts are $\fspin(6,1)\oplus \fsu(2)$ and $\fspin(5,2)\oplus
\fsu(2)$ respectively. $\ff_4$ has a non chiral odd part of type
(1,1). They are the proper de Sitter and anti de Sitter
superalgebras. For $d\leq 7$ one can find a simple superalgebra
whose bosonic part has the de Sitter or anti de Sitter algebra as
a factor. For higher dimensions this is not true, as one can check
directly from Table \ref{desitter}.

The contractions of these algebras to Poincar\'e superalgebras
where studied in detail in Ref.\cite{dflv} for $N=1$.  It is of
interest  to note that the contractions give super Poincar\'e
algebras with  a number of odd generators which is not, in
general,  the minimal one in super Poincar\'e. In Table
\ref{desitter} we have added the dimension of the odd part of the
superalgebra (``odd") together with the dimension of the odd part
of the $N$-extended super Poincar\'e algebra for a spacetime of
type $(d-1,1)$ (``odd SP"). The de Sitter superalgebra gives the
correct contraction for $d=4,5,12$. The anti de Sitter
superalgebra gives the correct contraction for $d=4,5,7,11,12$.
The remaining case give twice the number of odd generators. For
$d=6,10$ one obtains by contraction a non chiral algebra (In both,
de Sitter and anti de Sitter), and the same happens if one makes
the contraction of $\ff_4$. Physically, in dimension $d\leq 7$
supergravity theories exist with both, Minkowski and anti de
Sitter supersymmetric solutions.

\section{Examples}

We consider some examples in  $d=10$ and $d=4$.

\paragraph{ Spacetime of type (9,1).} We have $d=2$ mod 8 and
$\rho=0$ mod 8. The chiral spinor modules $S^\pm$ are real of
dimension $2n=16$. There is a chiral Poincar\'e superalgebra. The
$N=2$ chiral super Poincar\'e algebra is the IIB algebra. One can
also construct a non chiral one with the odd generators in the
direct sum $S^+\oplus S^-$ and it is the IIA algebra. Both have 32
odd spinor charges.

The conformal algebra is $\fso(10,2)$. Poincar\'e superalgebras
are subalgebras of the conformal superalgebras. The chiral algebra
of type (1,0) is a subalgebra of  $\fosp(1|32)$, the embedding
given by $$\begin{CD}\fsp(32)@<\supset<<\fso(10,2)
\\
 {\bf 32}@>>> S^+={\bf 32}
\end{CD}$$

 Type IIB
algebra (type (2,0)) is a subalgebra of $\fosp(2-q,q|32)$, with
$q=0,1$ and $q=0$ for compact R-symmetry. It has  64 spinor
charges. Finally, this superalgebra is embedded (not as a
subalgebra) into another superalgebra with 64 spinor charges,
$\fosp(1|64)$, which has the interpretation of the superconformal
algebra of a spacetime of type $(10,1)$ (that is, one dimension
more).

 The embedding of the even parts and
decompositions of the representations which are the odd part are
as follows

$$\begin{CD}\fsp(64)@<\supset<<\fso(2)\oplus \fsp(32)
@<\supset<<\fso(2)\oplus\fso(10,2)
\\
 {\bf 64}@>>>({\bf 2},{\bf 32})@>>>({\bf 2}, S^+)=({\bf 2},{\bf 32})
\end{CD}$$

Type IIA is also embedded into a superconformal algebra with 64
spinor charges, $\fosp(1,N2n)=\fosp(1|64)$. As before, we have

$$\begin{CD}\fsp(64)@<\supset<<\fspin(11,2)@<\supset<<\fso(10,2)
\\
 {\bf 64}@>>>S={\bf 64}@>>>S^+\oplus S^-={\bf 32}^+ \oplus {\bf 32}^-.
\end{CD}$$

\paragraph{The $d=4$ case}

It is interesting to consider the case of dimension 4 with all
possible signatures $\rho=0,2,4$. For $\rho=4$ (Euclidean case)
the superalgebra is $\fsu^*(4|2N)$, for $\rho=2$ (Lorentzian case)
the superalgebra is $\fsu(2,2|N)$ and for $\rho=0$ it is
$\fsl(4|N)$.

The superalgebras with 32 charges are $\fsu^*(4|4)$, $\fsu(2,2|4)$
and $\fsl(4|4)$ (since $n=m$ these algebras have no $\fu(1)$ or
$\fo(1,1)$ factor). They correspond to the underlying symmetries
of $N=4$ Euclidean Yang-Mills \cite{mc,bvv}, $N=4$ ordinary
Yang-Mills and $N=4$ self dual Yang-Mills \cite{si} considered in
the literature. Only the latter two exist with eight charges,
corresponding to $N=1$ supersymmetry, $\fsu(2,2|1)$ and
$\fsl(4|1)$.

We note that these minimal superconformal algebras have a further
extension into $\fosp(1|8)$ \cite{dflv,vv}, since $\fsp(8,\R)$
contains both, $\fsu(2,2)\oplus\fu(1)$ and $\fsl(4,\R)\oplus\R$.
$\fosp(1|8)$ can also be viewed as an anti de Sitter super algebra
in $d=5$, so by contraction we get the five dimensional super
Poincar\'e algebra with $Z$-charges. The $Z$-charges do not appear
if we make the contraction from the minimal superalgebra
$\fsu(2,2|1)$.

It is interesting to note that the enlargement of
$\fsu(2,2)\oplus\fu(1)$ to $\fsp(8,\R)$ does not change the rank
of the algebra, so the number of  quantum numbers that label an
irreducible unitary representation would be the same.

$\fosp(1|8)$ has (as all superconformal algebras) an $\fo(1,1)$
grading
$$\fosp(1|8)=\La^{-1}\oplus\Qa^{-1/2}\oplus\La^{0}\oplus\Qa^{+1/2}\oplus\La^{+1},$$
with $\La^0=\fsl(4,\R)\oplus\fso(1,1)$. Note that
$\fsl(4,\R)=\fspin(3,3)$ and that we have
$$\fso(3,1)\oplus\fso(2)\in \fso(3,3), \qquad (\rho=2)$$ with
$\fso(2)$ being the R-symmetry of $\fsu(2,2|1)$ and
$$\fso(2,2)\oplus\fso(1,1)\in \fso(3,3), \qquad (\rho=0)$$ with
$\fso(1,1)$ being the R-symmetry of $\fsl(4|1)$.

\section*{Appendix}
In this appendix we give some definitions that are used throughout
the paper. There are many references where these concepts are
treated in great detail. We cite here the ones we have used
\cite{be,ko,ka,le}.

\subparagraph{A.} A {\it super Lie algebra} is a $\Z_2$-graded
vector space $\fg=\fg_o+\fg_1$ with a bilinear operation
$[\;,\;]:\fg\times\fg\rightarrow\fg$ satisfying the following
properties:

\smallskip

\noindent {\bf a.} We say that an element $a\in \fg$ is
homogeneous of degree $p_a=0,1$ if $a\in \fg_0$ (a is even) or
$a\in \fg_1$ (a is odd) respectively. We then have that for $a$
and $b$ homogeneous elements of $\fg$ $$p_{[a,b]}=p_a+p_b\qquad
\hbox{modulo } \Z_2.$$

\smallskip

\noindent {\bf b.} The bracket is graded-skew symmetric,
$$[a,b]=-(-1)^{p_a\cdot p_b}[b,a],$$ with $a$ and $b$ homogeneous
in $\fg$.

\smallskip

\noindent {\bf c.} Generalized Jacobi identity, $$(-1)^{p_a\cdot
p_c}[a,[b,c]]+(-1)^{p_c\cdot p_b}[c,[a,b]]+(-1)^{p_b\cdot
p_a}[b,[c,a]]=0$$ if $a$, $b$ and $c$ are homogeneous in $\fg$.

\smallskip

It follows that $\fg_0$ is an ordinary Lie algebra and that the
subspace of the odd elements $\fg_1$ carries a representation of
$\fg_0$.

\bigskip

A superalgebra $\fg$ is {\it simple} if it has no other ideal than
$0$ and $\fg$. If $\fg$ is simple, then the representation of
$\fg_0$ on $\fg_1$ is faithful and $[\fg_1,\fg_1]=\fg_0$.  If
these two conditions are satisfied and, in addition the
representation of $\fg_0$ on $\fg_1$ is irreducible, then $\fg$ is
simple.

\subparagraph{B.} We give here the definition of some classical
complex superalgebras that are used in the text.

\bigskip

Let $V=V_0\oplus V_1$ be a $\Z_2$ graded vector space over $\C$
with $\dim V_0=m$ and $\dim V_1=n$. Then we have that the
endomorphisms of $V$ are also a graded vector space. In terms of a
basis
\begin{eqnarray*}&\mathrm{End}(V)=\left\{\begin{pmatrix} A_{m\times
m}&B_{n\times m}\\C_{m\times n}&D_{n\times
n}\end{pmatrix}\right\},\\&
\mathrm{End}(V)_0=\left\{\begin{pmatrix} A_{m\times
m}&0\\0&D_{n\times n}\end{pmatrix}\right\}, \quad
\mathrm{End}(V_1)=\left\{\begin{pmatrix} 0&B_{n\times
m}\\C_{m\times n}&0\end{pmatrix}\right\}.\end{eqnarray*} it is
endowed with a super Lie algebra structure with bracket
$$[a,b]=ab-(-1)^{p_ap_b}ba.$$ This superalgebra is denoted by
$\fgl(m|n)$.  We define the {\it supertrace} of an element of
$\fgl(m,n)$ as $$\mathrm{str}\begin{pmatrix}
A&B\\C&D\end{pmatrix}=\mathrm{tr}(A)-\mathrm{tr}(B).$$ The
subspace of elements of $\fgl(m,n,\C)$ that have zero supertrace
is a sub-superalgebra denoted by $\fsl(m,n,\C)$. If $m\neq n$,
then $\fsl(m|n,\C)$ is a simple superalgebra. Its  even part is
$\fsl(m,\C)\oplus\fsl(n,\C)\oplus \C$.

In $\fsl(n,n)$ there is a one dimensional ideal $\fii$ generated
by the matrix $\id_{2n\times 2n}$. The algebra $\fsl(n|n)/\fii$ is
also simple. Its even part is $\fsl(n,\C)\oplus\fsl(n,\C)$.

\bigskip

Let $F$ be a non degenerate bilinear form on the graded vector
space $V$.We assume that it is graded symmetric, that is,
$F(a,b)=(-1)^{p_ap_b}F(b,a)$.  This means that the restriction to
$V_0$ is symmetric and the restriction to $V_1$ is antisymmetric.
We assume also that $F(a,b)=0$ if $a\in V_0$ and $b\in V_1$, so
$V_0$ and $V_1$ are orthogonal. Because of the non degeneracy, we
have that $\dim(V_1)$ must be an even number. In a certain basis
the bilinear for is given by a matrix
$$\begin{pmatrix}\id_{m\times m}&0\\0&\Omega_{2p\times
2p}\end{pmatrix}, \quad \Omega_{2p\times
2p}=\begin{pmatrix}0&I\\-I&0\end{pmatrix}.$$ The subspace of
$\fgl(m|2p,\C)$ which satisfies $$a^tF+Fa=0, \qquad
a^t=\begin{pmatrix} A^T&C^T\\-B^T&D^T\end{pmatrix}$$ ($T$ denotes
the usual transpose) is a simple Lie superalgebra whose even part
is $\fso(m, \C)\otimes\fsp(2p,\C)$. It is called the
ortosymplectic algebra, $\fosp(m|2p)$.

\subparagraph{C.} The complex Lie superagebras defined above have
real forms that are real simple Lie superalgebras. These real
forms are determined by the real form of the even part (see Refs.
\cite{vp,dflv,dfl}). We list here the ones that are of interest
for our paper. The notation that we use for the real forms of Lie
algebras is the standard one \cite{he}.

\bigskip

\noindent {\it Real forms of $\fsl(m|n, \C)$}.

\medskip

\noindent 1. $\fsl(m|n,\R)$, with even part
$\fsl(m,\R)\oplus\fsl(n,\R)\oplus \R$.

\noindent 2. $\fsu(m|n)$, with even part
$\fsu(m)\oplus\fsu(n)\oplus \fu(1)$.

\noindent 3. $\fsu(m,n|p,q)$, with even part
$\fsu(m,n)\oplus\fsu(p,q)\oplus \fu(1) $.

\noindent 4. $\fsu^*(m|n)$, with even part
$\fsu^*(m)\oplus\fsu^*(n)\oplus \fso(1,1)$.

\bigskip

\noindent {\it Real forms of $\fosp(m|n, \C)$ $(n=2p)$}.

\medskip

\noindent 1. $\fosp(m|n,\R)$, with even part
$\fso(m,\R)\oplus\fsp(n,\R)$.

\noindent 2. $\fosp(m,q|n)$, with even part
$\fso(m,q)\oplus\fsp(n)$.

\noindent 3. $\fosp(m^*|2s,2t)$, with even part
$\fso^*(m)\oplus\fusp(2s,2t)$.

\bigskip

Additionally, there are other simple Lie superalgebras that are
constructed by taking the complex Lie superalgebra and looking at
it as a real Lie superalgebra space of twice the dimension. Their
even parts correspond to the complex even parts taken as real Lie
algebras. We denote those by $\fsl(n|m,\C)_\R$ and
$\fosp(n|m,\C)_\R$.

\section*{Acknowledgements}

S. F. would like to thank the Dipartimento di Fisica, Politecnico
di Torino for its kind hospitality during the completion of this
work. The work of S. F. has been supported in part by the European
Commission RTN network HPRN-CT-2000-00131, (Laboratori Nazionali
di Frascati, INFN) and by the D.O.E. grant DE-FG03-91ER40662, Task
C. M. A. Ll. would like to thank the Department of Physics and
Astronomy of the University of California, Los Angeles for its
hospitality during the completion of this work.

\end{document}